\documentclass{article}
\usepackage{graphicx}
\usepackage{amsmath}
\usepackage{amsfonts}
\usepackage{amssymb}
\renewcommand\baselinestretch1
\begin{document}

\title{ }

\bigskip
\ \ \ \ \ \ \ \ \ \ \ \ \ \ \ \ \ \ \ \ \ \ \ \ \ \ \ \ \ \ \ \ \ \ \ \ \ \ \ \ \ \ \ \ \ \ \ \ \ \ \ \ \ \ \ \ \ \ \ \ \ \ \ \ \ \ \ $\;\;\;\;\;\;\;\;\;\;\;\;\;\;\;\;\;\;\;\;\;\;\;\;\;\;\;\;\;\;\;\;\;\;\;\;\;\;\;\;\;\;\;\;\;\;\;\;$%
\begin{tabular}
[c]{l}%
{\small Submitted to Phys. Rev. B}%
\end{tabular}

\begin{center}
{\LARGE Time - dependent Ginzburg - Landau approach and application to superconductivity}

\bigskip

J. A. Zagrodzinski and T. Nikiciuk

\textit{Institute of Physics, Polish Academy of Sciences, 02-668 Warsaw, Poland}

email: zagro@ifpan.edu.pl and niki@ifpan.edu.pl

\bigskip
\end{center}

\textbf{Abstract}

A time dependent generalization of the Ginzburg -Landau Lagrangian is
proposed. It contains two terms determining the time dependence and the four
arbitrary scalar functions. Relevant equations, which coincide with equations
following from the suitable Hamiltonian, are derived by a standard variational
technique. These equations determine the energy conservation law and admit
twofold time dependence which leads either to first or to second order time
derivatives in Ginzburg - Landau equations. By introducing the gauge invariant
potentials and choosing the gauge which differs slightly from the classical
Lorentz one, the theory simplifies significantly. The results gained are
discussed and compared to some earlier propositions. The presented approach,
when reduced to a static one, is found to be in perfect agreement with that
reported recently by the Kol\'{a}\v{c}ek group. This indicates indirectly,
that the equation with the first order time derivative seems to be more justified.


\section{Introduction}

The Ginzburg - Landau (GL) free energy and the derived then consequently GL
equations have proven to be very fruitful in the theory of superconductivity.
Different situations, a lot of extensions, generalizations and applications to
the different type of superconductors can be found in a huge and commonly
known literature (c.f. \cite{Scht} , \cite{VDT}). But, perhaps as a natural
consequence of the original article \cite{GL}, the majority of papers and
textbooks starts from a static version, when time is completely ignored.

\textit{\ }The aim of the present paper is to correct this deficiency and to
incorporate time in the frame of GL formalism, so that the obtained relations
can be useful in the analysis of dynamical problems in the theory of
superconductivity. The static reduction should then coincide with the known
relations, issuing from the free energy concept.

There were different attempts to include time in the GL formalism. These
attempts can be divided into two groups. In the papers of the first group,
usually the time dependent formulation is postulated, starting a priori from
the time-dependent Ginzburg - Landau (TDGL) equation $\gamma\hslash\psi
_{t}=\left[  \left(  2m\right)  ^{-1}\left(  \hslash\nabla\right)  ^{2}%
+\alpha-\beta\left|  \psi\right|  ^{2}\right]  \psi,$ as in \cite{Sch} or
similar. Its applicability to superconductive problems is later discussed and
investigated. We are aware of a rich literature devoted to the mathematical
aspects of this equation and with a proper choice of constants with its
affinity to the nonlinear Schr\"{o}dinger equation, see e.g. \cite{CM}. The
second group, far less numerous, is formed by publications, in which the
authors start by choosing some reasonable Lagrangian, as e.g.\cite{Si},
\cite{KL1}. There are also some papers, in which time is introduced from other
premises e.g. \cite{Ri}.

We would like to point out that in the time dependence is involved also the
problem of the scalar potential and charge density. Can the gauge be chosen
such that scalar potential can vanish, and what is the role of a nonzero
charge density then? In many textbooks considering static consequences of GL
equations, the scalar potential is assumed to be zero without convincing
arguments. Better insight into this problem can gained from references
\cite{LKM}, \cite{KL1}, and references therein listing earlier contributions.
In this paper we propose a modification of the GL formalism in order to
include the time dependence starting from a suitable Lagrangian. Since the
authors are convinced that there may be different approaches to the problem,
in the first part there is proposed a general form of a GL-like Lagrangian, in
which time is included in two different manners. Moreover, there are included
four arbitrary scalar functions, which give additional freedom and assure some
universality of considerations. In principle, the whole approach is gauge
invariant. Next, the relevant modified GL equations are derived. These,
however, are rather complicated and illegible, when as independent quantities
are chosen the standard order parameter and potentials. Nevertheless, the
formalism can be considerably simplified by the introduction of so called
gauge - invariant (g.i.) potentials. This concept is only partially new, since
the first of these potentials is the commonly known gauge invariant phase.

The usefulness of this approach for superconductivity is discussed in the next
sections. We show that our treatment in the static case is in agreement with
the static approach reported in references\cite{KL1}, \cite{LKM},
\cite{KLB2},. More precisely, some particular static reduction of our general
theory is in very good agreement with the references cited, with an accuracy
to a single term, which exists in our theory, but probably was simplified in
the cited papers. This result serves as an important hint concerning the form
of TDGL. The discussed agreement indicates that the quantity, which has
physical interpretation as charge density must be independent of scalar
potential. A natural consequence is such reduction of TDGL\ that leads to the
presence of the first order time derivative in GL equation \cite{ZN1}, which
however seems to be in contrast to the case considered in reference \cite{Si}.

This last result automatically determines the final form of the TDGL-like
Lagrangian and, as a consequence, the TDGL\ equations.

\section{Time dependent Ginzburg - Landau approach}

We propose the extended version of the Lagrangian density which includes a
time dependence in the form%

\begin{equation}
\mathcal{L}=\mathcal{L}_{GL}+\mathcal{L}_{T}, \label{b1}%
\end{equation}%
\begin{align}
\mathcal{L}_{GL}  &  =\frac{1}{2}\left[  \left(  \mathbf{A}_{t}+\nabla
\varphi\right)  ^{2}-\left(  \nabla\times\mathbf{A}\right)  ^{2}\right]
-W-V\left[  \left(  -i\nabla-\mathbf{A}\right)  \psi\right]  \left[  \left(
i\nabla-\mathbf{A}\right)  \tilde{\psi}\right]  ,\label{b2}\\
\mathcal{L}_{T}  &  =M\left[  i\left(  \tilde{\psi}\partial_{t}\psi
-\psi\partial_{t}\tilde{\psi}\right)  -2\psi\tilde{\psi}\varphi\right]
+N\left[  \left(  i\partial_{t}-\varphi\right)  \psi\right]  \left[  \left(
-i\partial_{t}-\varphi\right)  \tilde{\psi}\right]  . \label{b3}%
\end{align}

The canonical quantities here are: vector and scalar potentials $\mathbf{A}$
and $\varphi,$ complex order parameters $\psi$ and $\tilde{\psi},$ and the
density of the charge carriers $n.$ We assume that the functions
$W,V,M,N\;$are algebraic real functions of two arguments: $\left|
\psi\right|  \;$and $n,\;$i.e. $W=W\left(  q,n\right)  $ with$.q:=\left|
\psi\right|  ,\;$etc$.$ and play a role of weight functions which can be fixed
for concrete problems. Later we will impose the symmetry condition that
$\tilde{\psi}=\psi^{\ast}.$

Choosing e.g. $W=\left|  \psi\right|  ^{2}+2\left|  \psi\right|  ^{4}$ and
dropping the term $\left(  \mathbf{A}_{t}+\nabla\varphi\right)  ^{2},$ it is
seen that $\mathcal{L}_{GL}$ coincides with the standard, static version of GL
free energy which can be found elsewhere, (e.g. \cite{Scht}, \cite{VDT} ).
Some particular choices of $W,V,M,N$ were reported in \cite{SS} and \cite{Si},
announced also in\cite{ZN1} and in the static version discussed in numerous
books and papers e.g. \cite{Scht} \cite{GL}, \cite{KLB2}, .\ Thus the
$\mathcal{L}_{GL}$ describes the static part of the Lagrangian density\ and
$\mathcal{L}_{T},$ the supplement introducing a twofold time dependence in
further field equations. We introduce canonical variable $n$ describing the
density of free charge carriers under influence of the lecture and paper
\cite{KLB2}. Of course all these quantities depend on space coordinates
$x,y,z$ and time $t.$

The proposed Lagrangian density is gauge invariant with respect to the
transformation
\begin{equation}
\{\mathbf{A\Rightarrow A}-\nabla\chi,\;\varphi\Rightarrow\varphi+\chi
_{t},\;\psi\Rightarrow\psi\exp\left(  i\chi\right)  ,\;\tilde{\psi}%
\Rightarrow\tilde{\psi}\exp\left(  -i\chi\right)  ,\;\theta\Rightarrow
\theta+\chi,\;n\Rightarrow n\}, \label{b4}%
\end{equation}
which will be evident shortly below (under assumption $\tilde{\psi}=\psi
^{\ast})$.

The standard variational procedure leads to the system of rather complicated equations%

\begin{align}
\delta_{\mathbf{A}}  &  :\left(  \mathbf{A}_{t}+\nabla\varphi\right)
_{t}+\nabla\times\nabla\times\mathbf{A}+V\left[  i\left(  \tilde{\psi}%
\nabla\psi-\psi\nabla\tilde{\psi}\right)  +2\mathbf{A}\psi\tilde{\psi}\right]
=0\label{b11}\\
\delta_{\varphi}  &  :-\Delta\varphi-\nabla\left(  \mathbf{A}_{t}\right)
-2M\psi\tilde{\psi}+N\left[  2\varphi\psi\tilde{\psi}+i\left(  \psi\tilde
{\psi}_{t}-\tilde{\psi}\psi_{t}\right)  \right]  =0\label{b12}\\
\delta_{n}  &  :\left\{
\begin{array}
[c]{c}%
W_{n}+V_{n}\left[  \left(  -i\nabla-\mathbf{A}\right)  \psi\right]  \left[
\left(  i\nabla-\mathbf{A}\right)  \tilde{\psi}\right]  -M_{n}\left[  i\left(
\tilde{\psi}\partial_{t}\psi-\psi\partial_{t}\tilde{\psi}\right)  -2\psi
\tilde{\psi}\varphi\right] \\
-N_{n}\left[  \left(  i\partial_{t}-\varphi\right)  \psi\right]  \left[
\left(  -i\partial_{t}-\varphi\right)  \tilde{\psi}\right]  =0
\end{array}
\right. \label{b13}\\
\delta_{\tilde{\psi}}  &  :\left\{
\begin{array}
[c]{c}%
W_{\tilde{\psi}}-V_{\tilde{\psi}}\left[  \left(  i\nabla+\mathbf{A}\right)
\psi\right]  \left[  \left(  i\nabla-\mathbf{A}\right)  \tilde{\psi}\right]
-M_{\tilde{\psi}}\left[  i\left(  \tilde{\psi}\partial_{t}\psi-\psi
\partial_{t}\tilde{\psi}\right)  -2\psi\tilde{\psi}\varphi\right] \\
+N_{\tilde{\psi}}\left[  \left(  i\partial_{t}-\varphi\right)  \psi\right]
\left[  \left(  i\partial_{t}+\varphi\right)  \tilde{\psi}\right]  +V\left[
-\Delta\psi+i\left(  \psi\nabla\cdot\mathbf{A}+2\mathbf{A\cdot}\nabla
\psi\right)  +\mathbf{A}^{2}\psi\right] \\
+N\left[  \partial_{t}^{2}\psi+i\left(  \psi\partial_{t}\varphi+2\varphi
\partial_{t}\psi\right)  -\varphi^{2}\psi\right]  +2M\left[  \psi
\varphi-i\partial_{t}\psi\right] \\
-\left(  \nabla\psi-i\mathbf{A}\psi\right)  \nabla V+\left(  \partial_{t}%
\psi+i\varphi\psi\right)  \partial_{t}N-iq\partial_{t}M=0
\end{array}
\right.  \label{b14}%
\end{align}

Equations (\ref{b11}) and (\ref{b12}) can be rewritten as nonhomogeneous wave
equations for potentials%

\begin{align}
\Delta\mathbf{A}-\mathbf{A}_{tt}  &  =V\left[  i\left(  \tilde{\psi}\nabla
\psi-\psi\nabla\tilde{\psi}\right)  +2\mathbf{A}\psi\tilde{\psi}\right]
+\nabla\left(  \nabla\cdot\mathbf{A}+\varphi_{t}\right) \label{b16}\\
\Delta\varphi-\varphi_{tt}  &  =-2M\psi\tilde{\psi}+N\left[  2\varphi
\psi\tilde{\psi}+i\left(  \psi\tilde{\psi}_{t}-\tilde{\psi}\psi_{t}\right)
\right]  -\left(  \nabla\cdot\mathbf{A}+\varphi_{t}\right)  _{t} \label{b17}%
\end{align}

The last terms in both equations vanish, if the Lorentz gauge $\nabla
\cdot\mathbf{A}+\varphi_{t}=0$ is assumed. Then the first terms on the right
hand sides represent the current $\mathbf{j}$ and charge $\rho$ densities,
respectively. Thus
\begin{align}
\mathbf{j}  &  =V\left[  i\left(  \tilde{\psi}\nabla\psi-\psi\nabla\tilde
{\psi}\right)  +2\mathbf{A}\psi\tilde{\psi}\right] \label{b18}\\
\rho &  =2M\psi\tilde{\psi}-N\left[  2\varphi\psi\tilde{\psi}+i\left(
\psi\tilde{\psi}_{t}-\tilde{\psi}\psi_{t}\right)  \right]  \label{b19}%
\end{align}

Assuming $\tilde{\psi}=\psi^{\ast}$ and calculating the imaginary part of
$\left[  (\text{\ref{b14}})\;\exp\left(  -i\theta\right)  \right]  $ one
obtains immediately the continuity equation $\nabla\cdot\mathbf{j}+\rho_{t}=0
$ with $\mathbf{j}$ and $\rho$ given by (\ref{b18}) and (\ref{b19}), respectively.

Thus in the presented formalism we are able to include time dependence and
four different types of weight functions. The physical interpretation will be
evident later on. Here, it is worthwhile to remark that a uniform and time
independent charge density of the background or lattice can be introduced by a
suitable choice of $M$ function.

As it was mentioned above, all derived relations are gauge invariant with
respect to (\ref{b4}), although it is not seen directly. It can be done quite
immediately defining new variables $\mathbf{F}$ and $f$ which surely are gauge
invariant by relations
\begin{equation}
\mathbf{F:=}\nabla\theta-\mathbf{A,\;\;\;\;\;}f:=\varphi+\theta_{t}.
\label{b20}%
\end{equation}
It is obvious that the first quantity $\mathbf{F}$ is the gauge invariant
phase commonly introduced description different superconductive problems. The
second one, $f$ was to our knowledge never defined. Both quantities redefine
the vector and scalar potentials and therefore we shall call them the gauge
invariant potentials (g.i. potentials).

It appears that one can completely rewrite the Lagrangian density (\ref{b1} -
\ref{b3}) using g.i. potentials according to (\ref{b20}). We have then%

\begin{equation}
\mathcal{L}=\frac{1}{2}\left[  \left(  \mathbf{F}_{t}-\nabla f\right)
^{2}-\left(  \nabla\times\mathbf{F}\right)  ^{2}\right]  -W-\left[
q^{2}\mathbf{F}^{2}+\left(  \nabla q\right)  ^{2}\right]  V+N\left(
q^{2}f^{2}+q_{t}^{\;2}\right)  -2Mq^{2}f, \label{b51}%
\end{equation}
where $q^{2}=\psi\tilde{\psi}$ and the remaining symbols have the meaning as before.

Lagrangian density (\ref{b1}) - (\ref{b3}) is now automatically gauge
invariant since it is defined by the gauge invariant quantities. Considering
now $\mathbf{F,}f,q,n$ as new variables we obtain the system of equations%

\begin{align}
\delta_{\mathbf{F}}  &  :\left(  \mathbf{F}_{t}-\nabla f\right)  _{t}%
+\nabla\times\nabla\times\mathbf{F}+2q^{2}V\mathbf{F}=0,\label{b31}\\
\delta_{f}  &  :\Delta f-\nabla\cdot\mathbf{F}_{t}+2q^{2}\left(  M-Nf\right)
=0,\label{b32}\\
\delta_{n}  &  :W_{n}+V_{n}\left[  q^{2}\mathbf{F}^{2}+\left(  \nabla
q\right)  ^{2}\right]  -N_{n}\left(  q^{2}f^{2}+q_{t}^{\;2}\right)
+2q^{2}M_{n}f=0,\label{b33}\\
\delta_{q}  &  :\left\{
\begin{array}
[c]{c}%
V\Delta q-Nq_{tt}-\frac{1}{2}W_{q}-q\left(  \mathbf{F}^{2}V-f^{2}N\right)
-\frac{1}{2}\left[  \mathbf{F}^{2}q^{2}-\left(  \nabla q\right)  ^{2}\right]
V_{q}-N_{n}n_{t}q_{t}\\
+\frac{1}{2}\left[  q^{2}f^{2}-\left(  q_{t}\right)  ^{2}\right]  N_{q}%
+V_{n}\left[  \nabla n\cdot\nabla q\right]  -fq\left(  2M+qM_{q}\right)  =0
\end{array}
\right.  \label{b34}%
\end{align}

Rewriting the first two equations as before, we have the nonhomogeneous wave equations%

\begin{align}
\Delta\mathbf{F-F}_{tt}  &  =2q^{2}V\mathbf{F+}\nabla\left(  \nabla
\cdot\mathbf{F-}f_{t}\right)  ,\label{b35}\\
\Delta f-f_{tt}  &  =-2q^{2}\left(  M-Nf\right)  +\left(  \nabla
\cdot\mathbf{F}-f_{t}\right)  _{t}, \label{b36}%
\end{align}
with the current and charge density defined as
\begin{align}
\mathbf{j}  &  =2qV\mathbf{F},\label{b37}\\
\rho &  =2q^{2}\left(  M-Nf\right)  . \label{b38}%
\end{align}

Now it is easy to check the continuity equations $\nabla\cdot\mathbf{j+}%
\rho_{t}=0$. Indeed, by (\ref{b31}) and (\ref{b32})
\[
\nabla\cdot\mathbf{j}=\nabla\cdot\left(  2qV\mathbf{F}\right)  =-\nabla
\mathbf{F}_{tt}+\Delta f_{t}=-2\left[  q^{2}\left(  M-Nf\right)  \right]
_{t}=-\rho_{t}.
\]

Two possible charge density components are given by (\ref{b38}) and it is only
a question of a concrete physical problem which of them can be neglected,
putting the relevant constant equal to zero. (The are no inevitable arguments
to assume $N=0;$ \ e.g. in paper \cite{KLB1} there is considered the
configuration when $\rho\sim\varphi.)$

As in the case where on the vector and scalar potentials the Lorentz gauge
could be imposed, now it can be also done formulating the Lorentz-like gauge
imposed on g.i. potentials in form of condition
\begin{equation}
\nabla\cdot\mathbf{F}-f_{t}=0. \label{b39}%
\end{equation}

This requirement itself is also gauge invariant since it contains gauge
invariant quantities, in contrast to the Lorentz gauge condition for
$\mathbf{A} $ and $f$ , which is not invariant with respect gauge (\ref{b4}),
unless $\Delta\theta-\theta_{tt}=0.$ It is obvious, since in the language of
standard potentials equation (\ref{b39}) reduces to $\nabla\cdot A+\varphi
_{t}=\Delta\theta-\theta_{tt}.$

When the static solutions are considered $f\equiv\varphi$ and the system of
equations (\ref{b31}) - (\ref{b39}) reduces to%

\begin{align}
&  \Delta\mathbf{F}-2q^{2}V\mathbf{F}=0,\label{b41}\\
&  \Delta f+2q^{2}\left(  M-Nf\right)  =0,\label{b42}\\
&  W_{n}+V_{n}\left[  q^{2}\mathbf{F}^{2}+\left(  \nabla q\right)
^{2}\right]  -N_{n}q^{2}f^{2}+2q^{2}M_{n}f=0,\label{b43}\\
&
\begin{array}
[c]{c}%
V\Delta q-\frac{1}{2}\left[  W_{q}+\left(  \mathbf{F}^{2}q^{2}-\left(  \nabla
q\right)  ^{2}\right)  V_{q}-q^{2}f^{2}N_{q}\right]  -q\left(  \mathbf{F}%
^{2}V-f^{2}N\right)  +V_{n}\left[  \nabla n\cdot\nabla q\right] \\
\;\;\;\;\;\;\;\;\;\;\;\;\;\;\;\;\;\;\;\;\;\;\;\;\;\;\;\;\;\;\;\;\;\;\;\;\;\;\;\;\;\;\;\;\;\;\;\;\;\;\;\;\;\;\;\;=fq\left(
2M+qM_{q}\right)  ,
\end{array}
\label{b44}%
\end{align}
where we assumed the static reduction of the Lorentz gauge for q.i. potentials
$\nabla\cdot\mathbf{F=0.}$ The current and charge density definitions
(\ref{b31}) and (\ref{b32}), respectively, remain unchanged.

Defining the canonical momenta $\mathbf{p}_{F},p_{f},p_{q},p_{n}\;$associated
with variables \ $\mathbf{F,}f,q,n$ in a standard form, we have%

\begin{equation}
\mathbf{p}_{F}=\mathbf{F}_{t}-\nabla f,\;\;p_{f}=0,\;\;p_{q}=2Nq_{t}%
,\;\;p_{n}=0. \label{b50}%
\end{equation}
Therefore the Hamiltonian density $\mathcal{H}=\mathbf{F}_{t}\cdot
\mathbf{p}_{F}+q_{t}p_{q}\mathcal{-L\;\;}$is%

\begin{equation}
\mathcal{H}=\frac{1}{2}\left[  \mathbf{p}_{F}^{2}+\left(  \nabla
\times\mathbf{F}\right)  ^{2}\right]  +\nabla f\cdot\mathbf{p}_{F}%
+\frac{\left(  p_{q}\right)  ^{2}}{4N}+W+\left[  q^{2}\mathbf{F}^{2}+\left(
\nabla q\right)  ^{2}\right]  V+2Mq^{2}f-Nq^{2}f^{2} \label{b5o1}%
\end{equation}
where the whole term $\left(  p_{q}\right)  ^{2}/4N$ drops out if $N=0$ and
then also $p_{q}=0.$ The standard Hamilton equations reproduce (\ref{b50}) and
equations\ (\ref{b31} - \ref{b34}), as should be. Substituting (\ref{b50})
into (\ref{b51}), the energy density is%

\begin{align}
\mathcal{E}  &  =\frac{1}{2}\left[  \mathbf{F}_{t}^{2}-\left(  \nabla
f\right)  ^{2}+\left(  \nabla\times\mathbf{F}\right)  ^{2}\right]  +W+\left[
q^{2}\mathbf{F}^{2}+\left(  \nabla q\right)  ^{2}\right]  V+\label{b52}\\
&  +2Mq^{2}f+N\left[  \left(  q_{t}\right)  ^{2}-f^{2}\right]  .\nonumber
\end{align}

It is natural to ask about the conservation relations. Tedious, though simple
calculations lead to the equation%

\begin{align}
&  \partial_{t}\left[  \frac{1}{2}\left[  \left(  \mathbf{F}_{t}\right)
^{2}+\left(  rot\mathbf{F}\right)  ^{2}-\left(  \nabla f\right)  ^{2}\right]
+W+\left[  q^{2}F^{2}+\left(  \nabla q\right)  ^{2}\right]  V+2Mq^{2}%
f+N\left[  \left(  q_{t}\right)  ^{2}-f^{2}\right]  \right] \nonumber\\
&
\;\;\;\;\;\;\;\;\;\;\;\;\;\;\;\;\;\;\;\;\;\;\;\;\;\;\;\;\;\;\;\;\;\;\;\;\;=\nabla
\cdot\left(  \mathbf{F}_{t}\times rot\mathbf{F+}\left(  F_{t}-\nabla f\right)
f_{t}+2Vq_{t}\nabla q\right)  .\;\;\;{\LARGE \;} \label{b55}%
\end{align}

Till now, still the real functions $W,V,M,N$ are completely arbitrary and some
of them can vanish. Observe however, that the right hand side of (\ref{b55})
i.e. flux depends only on $V$. Moreover, meanwhile we did not impose the
Lorentz-like gauge (\ref{b39}) and also in particular cases the variable $n$
can vanish simplifying all previous relations.

Admitting however the Lorentz-like gauge (\ref{b39}) system of equations
(\ref{b31}) - (\ref{b36}) becomes%

\begin{align}
&  \Delta\mathbf{F}-\mathbf{F}_{tt}=j=2q^{2}V\mathbf{F,}\label{b61}\\
&  \Delta f-f_{tt}=-\rho=-2q^{2}\left(  M-Nf\right)  ,\label{b62}\\
&  W_{n}+V_{n}\left[  q^{2}\mathbf{F}^{2}+\left(  \nabla q\right)
^{2}\right]  -N_{n}\left(  q^{2}f^{2}+q_{t}^{\;2}\right)  +2q^{2}%
M_{n}f=0,\label{b63}\\
&  V\Delta q-Nq_{tt}=\frac{1}{2}W_{q}+q\left(  \mathbf{F}^{2}V-f^{2}N\right)
+\frac{1}{2}\left[  \mathbf{F}^{2}q^{2}-\left(  \nabla q\right)  ^{2}\right]
V_{q}\label{b64}\\
&  -\frac{1}{2}\left[  q^{2}f^{2}-\left(  q_{t}\right)  ^{2}\right]
N_{q}-V_{n}\left[  \nabla n\nabla q\right]  +N_{n}n_{t}q_{t}+fq\left(
2M+qM_{q}\right)  ,\nonumber
\end{align}

and in the next paragraph we shall discuss some particular situations due to
the specific choice of $W,V,M,\;$and $N\;$functions.

\section{Elementary reductions}

As a first example we consider the system of equations which follows from
Lagrangian (\ref{b1}) - (\ref{b3})\ and from the relevant equations when the
variable $n$ does not appear at all, $V=N=1$ and $M$ vanishes. Moreover,
assuming $W\left(  q\right)  =q^{2}+\beta q^{4}/2,$ from (\ref{b61}) -
(\ref{b64}) and (\ref{b55}) we obtain the system of equations:\ %

\begin{align}
\mathcal{L}  &  =\frac{1}{2}\left[  \left(  \mathbf{F}_{t}-\nabla f\right)
^{2}-\left(  \nabla\times\mathbf{F}\right)  ^{2}\right]  -W-\left[
q^{2}\mathbf{F}^{2}+\left(  \nabla q\right)  ^{2}\right]  +\left(  q^{2}%
f^{2}+q_{t}^{\;2}\right) \label{c1}\\
\Delta\mathbf{F}-\mathbf{F}_{tt}  &  =j=2q^{2}\mathbf{F\;}\label{c2}\\
\Delta f-f_{tt}  &  =-\rho=2q^{2}f,\label{c3}\\
\Delta q-q_{tt}  &  =\frac{1}{2}W_{q}+q\left(  F^{2}-f^{2}\right) \label{c4}\\
&  \partial_{t}\left[  \frac{1}{2}\left[  \left(  \mathbf{F}_{t}\right)
^{2}+\left(  rot\mathbf{F}\right)  ^{2}-\left(  \nabla f\right)  ^{2}\right]
+W+\left[  q^{2}\mathbf{F}^{2}+\left(  \nabla q\right)  ^{2}\right]  +\left[
\left(  q_{t}\right)  ^{2}-f^{2}\right]  \right] \nonumber\\
&  =\nabla\cdot\left(  \mathbf{F}_{t}\times rot\mathbf{F+}\left(  F_{t}-\nabla
f\right)  f_{t}+2q_{t}\nabla q\right)  \label{c5}%
\end{align}

One can find elsewhere Its static version, but when the time dependence in
included, it was discussed only in a few publications e.g. \cite{Si}
\cite{SS},. To our knowledge it has no broader application in
superconductivity, probably because of two rather serious insufficiencies: the
charge is proportional to the scalar potentials (see r.h.s. of (\ref{c3}) and
the second order derivatives with respect to time $q_{tt}$ in equation
(\ref{c4}) appears. Rewriting this system in a standard language of
$A,\varphi,\psi, $ we will have the second derivative $\psi_{tt},$ which seems
to be at least strange when the time dependent Ginzburg - Landau equation (or
Gross - Pitaevskiii eqn.) are compared with the classical Schr\"{o}dinger equation.

In order to derive the system of equations with the first order derivative
$\psi_{t}$ in relevant GL equation, as e.g. in paper \cite{Sch} one can
consider a dual situation when $V=M=1$ and $N\;$vanishes ( also $n\equiv0$).
\ The system (\ref{b61}) - (\ref{b64}) and (\ref{b55}) reduces then to%

\begin{align}
\mathcal{L}  &  =\frac{1}{2}\left[  \left(  \mathbf{F}_{t}-\nabla f\right)
^{2}-\left(  \nabla\times\mathbf{F}\right)  ^{2}\right]  -W-\left[
q^{2}\mathbf{F}^{2}+\left(  \nabla q\right)  ^{2}\right]  -2q^{2}%
f,\label{c11}\\
\Delta\mathbf{F}-\mathbf{F}_{tt}  &  =j=2q^{2}\mathbf{F,\;}\label{c12}\\
\Delta f-f_{tt}  &  =-\rho=-2q^{2},\label{c13}\\
\Delta q  &  =\frac{1}{2}W_{q}+q\mathbf{F}^{2}+2fq,\label{c14}\\
&  \partial_{t}\left[  \frac{1}{2}\left[  \left(  F_{t}\right)  ^{2}+\left(
rot\mathbf{F}\right)  ^{2}-\left(  \nabla f\right)  ^{2}\right]  +W+\left[
q^{2}\mathbf{F}^{2}+\left(  \nabla q\right)  ^{2}\right]  +2q^{2}f\right]
\nonumber\\
&  =\nabla\cdot\left(  \mathbf{F}_{t}\times rot\mathbf{F+}\left(
\mathbf{F}_{t}-\nabla f\right)  f_{t}+2q_{t}\nabla q\right)  . \label{c16}%
\end{align}

In contrast to the previous choice, now in language of g.i. potentials,
equation (\ref{c14}) does not contain the time derivative, but it is present
when the system is written by means of $\mathbf{A},\varphi,\psi$
variables.\ In fact, instead of (\ref{c14}), equation (\ref{b14}) reduces to form%

\begin{equation}
i\partial_{t}\psi=\frac{1}{2}\left(  i\nabla+\mathbf{A}\right)  ^{2}\psi
+\frac{1}{2}W_{\tilde{\psi}}+\psi\varphi. \label{c17}%
\end{equation}
A comparison of both above discussed examples speaks in favor of the choice
$N=0,$ since it leads to the first order in time GL equation.

In the next paragraph we consider a further argument in favor of that choice.

\section{Application to the superconductivity}

In a series of papers \cite{KL1} \cite{KLB2}, there was considered the Bardeen
phenomenological extension of the static GL theory, mainly in order to include
the energy of electric field and to demonstrate an important role of the
scalar potential which is very often neglected. Unfortunately this theory is
completely static, starting from the time independent free energy. Neglecting
the unimportant details and in symbols (signature) adopted in present paper
the free energy discussed there has the form%

\begin{equation}
\mathcal{L=}\frac{1}{2}\left[  -\left(  \nabla\varphi\right)  ^{2}+\left(
\nabla\times\mathbf{A}\right)  ^{2}\right]  +U+w+\frac{n}{4m}\left|  \left(
i\nabla+\mathbf{A}\right)  \psi\right|  ^{2}+\varphi\rho\label{d1}%
\end{equation}
There is introduced a new variable $n$ $-$ the density of the total charge,
while $\psi$ is reserved for the order parameter. The function $w$ includes
the Gorter-Casimir corrections and $w=-\left[  \frac{1}{4}T_{c}^{2}\left|
\psi\right|  ^{2}-\frac{1}{2}T^{2}\sqrt{1-\left|  \psi\right|  ^{2}}\right]
\gamma.$ Essential for further consideration is that $\gamma=\gamma\left(
n\right)  $ and thus $w=w\left(  \left|  \psi\right|  ,n\right)  .$ Moreover,
the effective mass of the carriers also depends on density of charge i.e.
$m=m\left(  n\right)  .$ The function $U$ introduced in (\ref{d1}) described
the effect of the screening on the Thomas - Fermi length and what is important
$U=U\left(  n\right)  \;$and thus $\partial U/\partial n=C\rho,$ where C is
constant. From the context of the paper it follows also that authors assume
$\rho=en.$

According to the authors, variations with respect $A,\varphi,\psi$ and $n$
variables lead to the system of equations%

\begin{align}
\nabla\times\nabla\times\mathbf{A}  &  =\nabla\times H=-\frac{n}%
{2m}\operatorname{Re}\left(  \tilde{\psi}\left(  i\nabla+\mathbf{A}\right)
\psi\right) \label{d2}\\
-\Delta\varphi &  =\rho\label{d3}\\
e\varphi &  =-C\rho-w_{n}-\frac{1}{4}\frac{1}{m}\left(  1-\frac{d\left(  \ln
m\right)  }{d\left(  \ln n\right)  }\right)  \left|  \left(  i\nabla
+\mathbf{A}\right)  \psi\right|  ^{2}\label{d4}\\
\frac{n}{4m}\left(  i\nabla+\mathbf{A}\right)  ^{2}\psi &  =-\psi w_{\left|
\psi\right|  ^{2}} \label{d5}%
\end{align}

Accepting without discussion the physical aspect and phenomenological
justification of the free energy according to expression (\ref{d1}) we shall
show that one can propose an expression for the Lagrangian which includes time
dependence and next to derive the self-consistent time dependent field
equations which quite completely reduce to the system (\ref{d2}) - (\ref{d5}),
when only static effects are considered. ''Quite completely'' means here that
- up to the terms $0\left(  \nabla\left(  n/m\right)  \right)  .$ The reasons
that the terms proportional to $\nabla\left(  n/m\right)  $ can be neglected
in equation\ (\ref{d5}) was not specified. A proposition of a proper
Lagrangian reduces in practice to a suitable choice of four functions $W,V,M$
and $N$ .in our considerations.

Let assume that
\begin{equation}
V=n/\left(  4m\right)  ,\;\;W=\alpha n^{2}/2+w,\;\;M=\alpha n/\left(  2\left|
\psi\right|  ^{2}\right)  ,\;N=0, \label{d9}%
\end{equation}
where still $m=m\left(  n\right)  ,$ $w=w\left(  \left|  \psi\right|
^{2},n\right)  $ and $\alpha$ is constant. Lagrangian (\ref{b1}) takes the form%

\begin{align}
\mathcal{L}  &  =\frac{1}{2}\left[  \left(  \mathbf{A}_{t}+\nabla
\varphi\right)  ^{2}-\left(  \nabla\times\mathbf{A}\right)  ^{2}\right]
-\left(  \alpha n^{2}/2+w\right)  -\frac{n}{4m}\left|  \left(  i\nabla
+\mathbf{A}\right)  \psi\right|  ^{2}\label{d11}\\
&  \;\;\;\;\;\;\;\;\;\;\;\;\;\;\;\;\;\;\;\;\;\;\;\;\;\;\;\;+\frac{\alpha
n}{2\left|  \psi\right|  ^{2}}\left[  i\left(  \tilde{\psi}\partial_{t}%
\psi-\psi\partial_{t}\tilde{\psi}\right)  -2\psi\tilde{\psi}\varphi\right]
.\nonumber
\end{align}
It differs from the free energy (\ref{d1}) practically only by time dependent
terms. Consequently, the proper field equations (\ref{b11}) - (\ref{b14}) are%

\begin{align}
\delta_{\mathbf{A}}  &  :\Delta\mathbf{A}-\mathbf{A}_{tt}-\frac{n}{4m}\left[
i\left(  \tilde{\psi}\nabla\psi-\psi\nabla\tilde{\psi}\right)  +2\mathbf{A}%
\psi\tilde{\psi}\right]  =\nabla\left(  \nabla\cdot\mathbf{A}+\varphi
_{t}\right)  ,\label{d12}\\
\delta_{\varphi}  &  :\Delta\varphi-\varphi_{tt}=-\alpha n-\left(
\nabla\mathbf{A}+\varphi_{t}\right)  _{t}\;,\label{d13}\\
\delta_{n}  &  :a\left(  n+\varphi\right)  +w_{n}+\frac{1}{4}\left(  \frac
{n}{m}\right)  _{n}\left|  \left(  i\nabla+\mathbf{A}\right)  \psi\right|
^{2}-\frac{\alpha}{2\left|  \psi\right|  ^{2}}i\left(  \tilde{\psi}%
\partial_{t}\psi-\psi\partial_{t}\tilde{\psi}\right)  =0,\label{d14}\\
\delta_{\tilde{\psi}}  &  :\left|  \psi\right|  ^{2}w_{\left|  \psi\right|
^{2}}+\frac{n}{4m}\tilde{\psi}\left(  i\nabla+\mathbf{A}\right)  ^{2}%
\psi-\frac{1}{4}\left(  \tilde{\psi}\nabla\psi-i\mathbf{A}\left|  \psi\right|
^{2}\right)  \nabla\left(  \frac{n}{m}\right)  -i\frac{\alpha}{2}n_{t}=0.
\label{d15}%
\end{align}

Assuming a Lorentz gauge $\nabla\mathbf{A}+\varphi_{t}=0$ and observing that
$\left(  n/m\right)  _{n}=\left[  1-d\left(  \ln m\right)  /d\left(  \ln
n\right)  \right]  /m,$ it is seen that static version of the system
(\ref{d12})\ - (\ref{d15}) is quite equivalent to system (\ref{d2}) -
(\ref{d5}). The difference is only because of the term $-\frac{1}{4}\left(
\tilde{\psi}\nabla\psi-i\mathbf{A}\left|  \psi\right|  ^{2}\right)
\nabla\left(  n/m\right)  $ in the equation (\ref{d15}), which in (\ref{d5})
was probably omitted or simplified.

We would underline that the time dependent system (\ref{d12})\ - (\ref{d15})
is self-consistent with all consequences which were mentioned in the previous
paragraphs. The ''prototype'' of the GL equation (\ref{d14}) contains only the
first order time derivative.

Rewriting all equations in the language of the gauge invariant potentials
$\mathbf{F}$ and $f$, the whole theory becomes much simpler, particularly when
the Lorentz-like gauge is adopted. Substituting (\ref{d9}) to (\ref{b51}) -
(\ref{b34}) we obtain Lagrangian%

\begin{equation}
\mathcal{L}=\frac{1}{2}\left[  \left(  \mathbf{F}_{t}-\nabla f\right)
^{2}-\left(  \nabla\times\mathbf{F}\right)  ^{2}\right]  -\frac{\alpha n^{2}%
}{2}-w-\frac{n}{4m}\left[  q^{2}\mathbf{F}^{2}+\left(  \nabla q\right)
^{2}\right]  -\alpha nf, \label{d21}%
\end{equation}
and relevant field equations%

\begin{align}
\delta_{\mathbf{F}}  &  :\Delta\mathbf{F}-\mathbf{F}_{tt}-\frac{n}{2m}%
q^{2}\mathbf{F}=-\nabla\left(  \nabla\cdot\mathbf{F}-f_{t}\right)
_{t},\label{d22}\\
\delta_{f}  &  :\Delta f-f_{tt}+\alpha n=\left(  \nabla\cdot\mathbf{F}%
-f_{t}\right)  _{t},\label{d23}\\
\delta_{n}  &  :\alpha n+w_{n}+\left(  \frac{n}{4m}\right)  _{n}\left[
q^{2}\mathbf{F}^{2}+\left(  \nabla q\right)  ^{2}\right]  +af=0,\label{d24}\\
\delta_{q}  &  :\frac{n}{m}\Delta q-2w_{q}-\frac{n}{m}qF^{2}+\left(  \frac
{n}{m}\right)  _{n}\nabla n\cdot\nabla q=0, \label{d25}%
\end{align}
together with a conservation law\ (Poyinting like theorem)%

\begin{align}
&  \partial_{t}\left[  \frac{1}{2}\left[  \left(  F_{t}\right)  ^{2}+\left(
rot\mathbf{F}\right)  ^{2}-\left(  \nabla f\right)  ^{2}\right]  +w+\frac
{n}{4m}\left[  q^{2}\mathbf{F}^{2}+\left(  \nabla q\right)  ^{2}\right]
+\frac{n\alpha}{2}\left(  n+2f\right)  \right] \nonumber\\
&  \;\;\;\;\;\;\;\;\;\;\;\;\;\;=\nabla\cdot\left(  \mathbf{F}_{t}\times
rot\mathbf{F+}\left(  \mathbf{F}_{t}-\nabla f\right)  f_{t}+\frac{n}{2m}%
q_{t}\nabla q\right)  .\;\;\;{\LARGE \;} \label{d27}%
\end{align}

It is seen that the system (\ref{d21}) - (\ref{d27}) is much simpler than
those expressed by the standard potentials, especially when the condition
$\nabla\cdot\mathbf{F}-f_{t}=0$ is imposed. Moreover, the time derivatives do
not appear at all in equations (\ref{d24}) and (\ref{d25}), suggesting to use
the static solutions as a first step when the perturbation technique will be
applied for the solution of a dynamic problem.

Repeating the calculations of this paragraph in case when $N\neq0$ and it is
e.g.$N=\beta n/2\left|  \psi\right|  ^{2},$ where $\beta$ is constant, the
static versions of equations (\ref{d12}) and (\ref{d15}) do not change but the
static version of equations (\ref{d13}) and (\ref{d14}) take a form%

\begin{equation}
\Delta\varphi=-n\left(  \alpha+\beta\varphi\right)  ;\;\;\;a\left(
n+\varphi\right)  +w_{n}+\frac{1}{4}\left(  \frac{n}{m}\right)  _{n}\left|
\left(  i\nabla+\mathbf{A}\right)  \psi\right|  ^{2}-\frac{\beta}{2}%
\varphi^{2}=0. \label{d28}%
\end{equation}

It is seen that the agreement of this system with the system (\ref{d2}) -
(\ref{d5}), which was proposed on a base of physical premises can be reached
only when $\beta=0,$ i.e. when in the starting Lagrangian (\ref{b1})
-(\ref{b3}) $N$ function vanishes. The proposed term, proportional to $M,$
will be the sole one which includes time dependence.

\section{Conclusion}

We proposed a general and extended formula for the Lagrangian in spirit of the
Ginzburg - Landau formalism, augmented by the time-dependent terms of two
types. Moreover, a few arbitrary weight scalar functions introduced in this
Lagrangian allow us to derive a family of the self-consistent time - dependent
extended Ginzburg - Landau equations which coincide with the relevant Hamilton
equations. Among them there are families with first and second time
derivatives of the order parameter functions. The whole theory and equations
can be simplified when it is rewritten by means of the gauge - invariant
potentials. As an example of the application, the static version of GL
equations reported in\ \cite{KL1}\ \cite{KLB2}, are derived and verified as
the reduction of a full dynamical theory.

This reproducibility seems to indicate that for the correct dynamic theory,
the scalar function $N$ discussed in the text should vanish. This means that
then only the first order time derivative will be present in the time
dependent version of GL formalism.

\textbf{Acknowledgements}

The authors thank E. Infeld and S. Lewandowski for numerous remarks and
comments. This work was supported by KBN 123.456.79/02 grant. The European ESF
Program\ -\ Vortex is also gratefully acknowledged.

\end{document}